# Comment on "Evidence of Abundant and Purifying Selection in Humans for Recently Acquired Regulatory Functions"

Nicolas Bray[1] and Lior Pachter[1,2,*]

[1] Department of Mathematics, University of California, Berkeley, CA 94720, USA.
[2] Departments of Molecular & Cell Biology and Computer Science & Electrical Engineering, University of California, Berkeley, CA 94720, USA.
*To whom correspondence should be addressed.

Ward and Kellis (Reports, September 5 2012) identify regulatory regions in the human genome exhibiting lineage-specific constraint and estimate the extent of purifying selection. There is no statistical rationale for the examples they highlight, and their estimates of the fraction of the genome under constraint are biased by arbitrary designations of completely constrained regions.

Ward and Kellis (1) combine population genomic data from the 1000 Genomes Project (2) with biochemical data from the ENCODE project (3) to look for signatures of human constraint in regulatory elements. Their analysis is based on measuring three different proxies for constraint: SNP density, heterozygosity and derived allele frequency. To identify specific classes of regulatory regions under constraint aggregated regions associated with specific gene ontology (GO) categories were tested for significance. The tests produced 55 GO categories (Table S5) out of which two were highlighted in the paper: those for retinal cone cell development and nerve growth factor receptor signaling. Despite the fact that the listed categories were required to pass a false discovery rate (FDR) threshold for both the heterozygosity and derived allele frequency (DAF) measures, it is statistically invalid to highlight any specific GO category. FDR control merely guarantees a low false discovery rate among the entries in the entire list. Moreover, there is no obvious explanation for why categories such as chromatin binding (which has a smaller DAF than nerve growth) or protein binding (with the smallest p-value) appear to be under purifying selection. In fact, retinal cone cell development and nerve growth factor are 33 and 34 out of the 55 listed GO categories when sorted by the DAF p-value (42 and 54 when sorted by heterozygosity p-value).

The main result of the paper is the estimate that in addition to the 5% of the human genome conserved across mammalian genomes, at least another 4% is subject to lineage-specific constraint. This result is based on adding up the estimates of constrained nucleotides from Table S6 (using the derived allele frequency measure). These were calculated using a statistic that is computed as follows: for each one of ten bins determined according to estimated background selection strength, and for every feature $F$, the average DAF value $D_F$ was rescaled to $PUC_F = (D_F - D_{CNDC})/(D_{NCNE} - D_{CNDC})$, where $D_{CNDC}$ and $D_{NCNE}$ are the bin-specific average DAFs of conserved non-degenerate coding regions and non-conserved non-ENCODE regions respectively. One problem with the statistic is that the non-conserved regions contain nucleotides not conserved in all mammals, which is not the same as nucleotides not conserved in any mammals. The latter would be needed in order to identify human specific constraint. Second, the statistic $PUC_F$ is used as a proxy for the proportion under constraint even though, as defined, it could be less than zero or greater than one. Indeed, in Table S6 there are four values among the confidence intervals for the estimated proportions using DAF that include values less than 0% or above 100%. Ward and Kellis are therefore proposing that some features might have a negative number of nucleotides under constraint. Moreover, while it is possible that $PUC_F$ might correlate with the true proportion of nucleotides under constraint, there is no argument provided in (1). Thus, while Ward and Kellis claim to have estimated the proportion of nucleotides under constraint, they have only computed a statistic named "proportion under constraint".